\documentclass[a4paper]{jpconf}
\usepackage{graphicx}
\begin{document}
\title{A Multi-APD readout for EL detectors}

\author{T. Lux$^1$, O. Ballester$^1$, J. Illa$^1$, G. Jover$^1$, C. Martin$^1$, J. Rico$^{1,2}$, F. Sanchez$^1$}

\address{$^1$ Institut de F\'{i}sica d\'{}Altes Energies (IFAE), Edifici Cn, Universitat Aut\`{o}noma de Barcelona, 08193 Bellaterra (Barcelona), Spain}
\address{$^2$ Instituci\'{o} Catalana de Recerca i Estudis Avan\c{c}ats (ICREA), 08010 Barcelona, Spain}

\ead{Thorsten.Lux@ifae.es}

\begin{abstract}
Detectors with an electroluminesence readout show an excellence performance in respect of energy resolution making them interesting for various applications as X-ray detection, double beta and dark matter experiments, Compton and $\gamma$ cameras, etc. In the following the study of a readout based on avalanche photo diodes to detect directly the VUV photons is presented. Results of measurements with 5 APDs in xenon at pressures between 1 and 1.65 bar are shown indicating that such a readout can provide excellent energy and a moderate position resolution.   
\end{abstract}

\section{Introduction}
\begin{figure}[h]
\begin{center}
\begin{minipage}{14pc}
\begin{center}
\includegraphics[width=14pc]{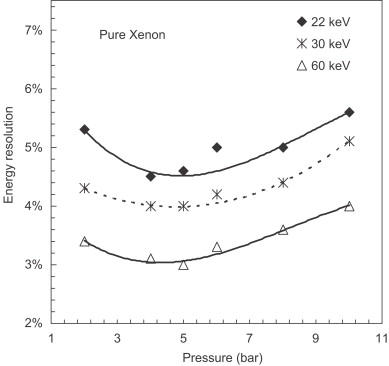}
\caption{\label{fig_coimbra1}Energy resolution and gain as function of the pressure for three different energies.\cite{Coelho2007444}}
\end{center}
\end{minipage}\hspace{2pc}%
\begin{minipage}{14pc}
\includegraphics[width=16pc]{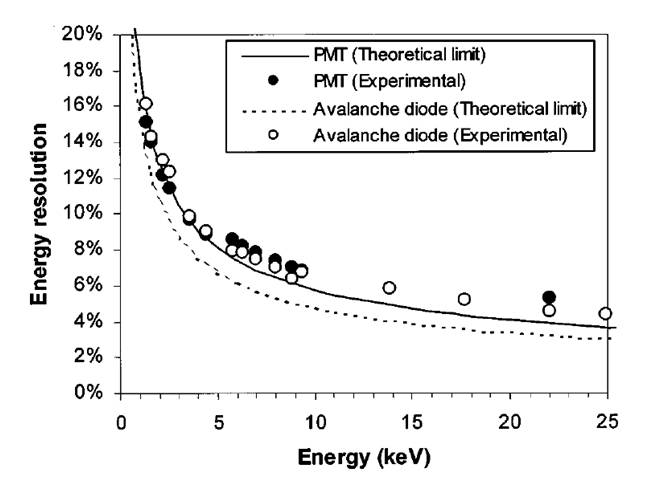}
\caption{\label{fig_coimbra2}Direct comparison for the achieved energy resolution with one PMT and one APD.\cite{Lopes2001gpsc}}
\end{minipage} 
\end{center}
\end{figure}

In conventional gas detectors the electrons released due to ionization within the active volume are drifted to the readout plane where a gas amplification takes place to amplify the primary signal before it is processed by the electronics. The disadvantage of this approach is that the exponential amplification introduces fluctuations in the measurement limiting the achievable energy resolution. In an electroluminsence (EL) readout this charge amplification process is replaced by a light amplification. Primary electrons drift to a region which is concluded by two wire meshes. Between the two meshes a voltage is applied such that the electrons excite the gas, normally pure noble gases as xenon or argon, without ionizing it. In the de-excitation of the atoms, photons in the VUV region are emitted isotropically. Since this is a linear process, the gain fluctuations introduced are smaller and therefore the achievable energy resolution is significantly better. The amount of photons produced depends on the applied voltage difference, the distance between the two meshes and also on the operation pressure. It turns out that the performance improves with increasing the pressure. In \cite{Monteiro:2007vz} a detailed description of the process for xenon can be found.\\
The double beta (DB) experiment NEXT ({\it Neutrino Experiment with a Xenon TPC}) is based on a xenon TPC operated at up to 10 bar. It is foreseen to use an EL readout to achieve the very good energy resolution of not more than 1 \% FWHM required at the {$Q_{\beta\beta}$} value (2.46 MeV). In the base design the chamber will be an asymmetric TPC in which the energy is measured by a PMT plane behind the cathode while the tracking information is obtained by a plane of MPPCs coated with TPB and covering about 1 \% of the total area. A detailed description of the base design can be found in \cite{Granena:2009it}. 
%Since MPPCs are known to be very temperature sensitive, having a large "`positive"' noise and a limited dynamic range which might deteroriate the tracking performance, the study of an alternative sensor seemed to be recomendable.\\
An alternative possibility using APDs to unify tracking and energy measurement in a single technology is discussed in this paper. The topic of the T$_0$ measurement in such a detector is not concerned here but will requiere additional R\&D studies. 
In the following, we present the performance study of an EL readout with 5 APDs as a first stage towards large readout areas. It turns out that such a readout based on APDs might not be only suitable for DB experiments like NEXT but also for other applications as e.g. Compton cameras.   
The use of APDs for the readout of EL chambers were already studied intensively by a research group at the University of Coimbra \cite{Coelho2007444}. Their results (fig.~\ref{fig_coimbra1}) indicate that an excellent energy resolution of less than 5 \% can be achieved at 22 keV for pressures between 4 to 6 bar. The authors believe that the rise for higher pressures is caused by experimental effects (micro sparks) in their setup. If this result could be scaled with $1/\sqrt{E}$ to the $Q_{\beta\beta}$ of xenon (2.46 MeV), an energy resolution of better than 0.5 \% FWHM at 2.46 MeV would be possible. They also compared directly the performance of reading out the chamber with an APD to the one achieved with a PMT \cite{Lopes2001gpsc}. The result is shown in fig.~\ref{fig_coimbra2}. As one can see the energy resolutions with these two kind of sensors are very similar under the same experimental conditions.
However, these measurements were performed with one single APD with a diameter of 16 mm (from Advanced Photonics Industries), while for the coverage of larger readout areas a multi-APD will be necessary.\\
We discuss here the result of a prototype to evaluate the performance of such readouts using an array of 5 APD sensors.
\section{Experimental Setup}
\subsection{The chamber}
\begin{figure}[h]
\begin{center}
\begin{minipage}{14pc}
\includegraphics[width=14pc]{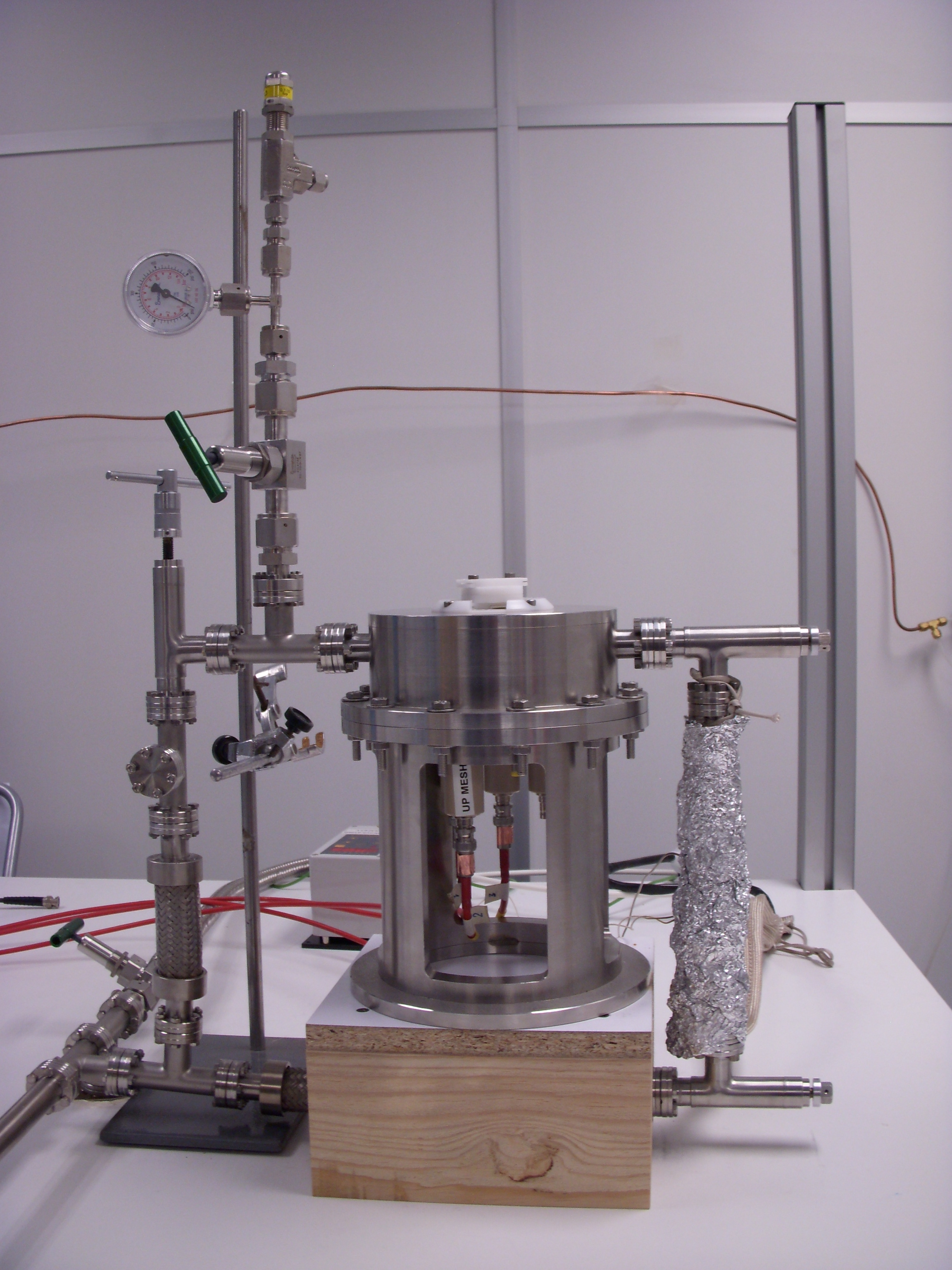}
\caption{\label{fig_chamber1}Photo of the chamber with the gas system loop.}
\end{minipage}\hspace{2pc}%
%\end{center}
%\end{figure}
%\begin{figure}[h]
%\begin{center}
\begin{minipage}{14pc}
\includegraphics[width=14pc]{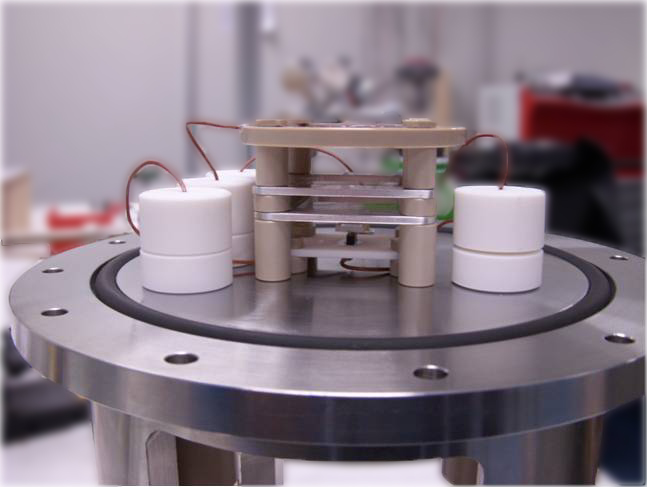}
\caption{\label{fig_chamber2}Side view of the inner tower. }
\end{minipage}
\end{center}
\end{figure}

\begin{figure}[h]
\begin{center}
\begin{minipage}{14pc}
\includegraphics[width=14pc]{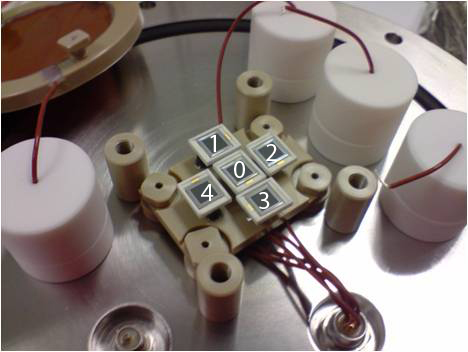}
\caption{\label{fig_chamber3}Photo of the 5 APDs with the numbering scheme.}
\end{minipage} \hspace{2pc}%
\begin{minipage}{14pc}
\includegraphics[width=14pc]{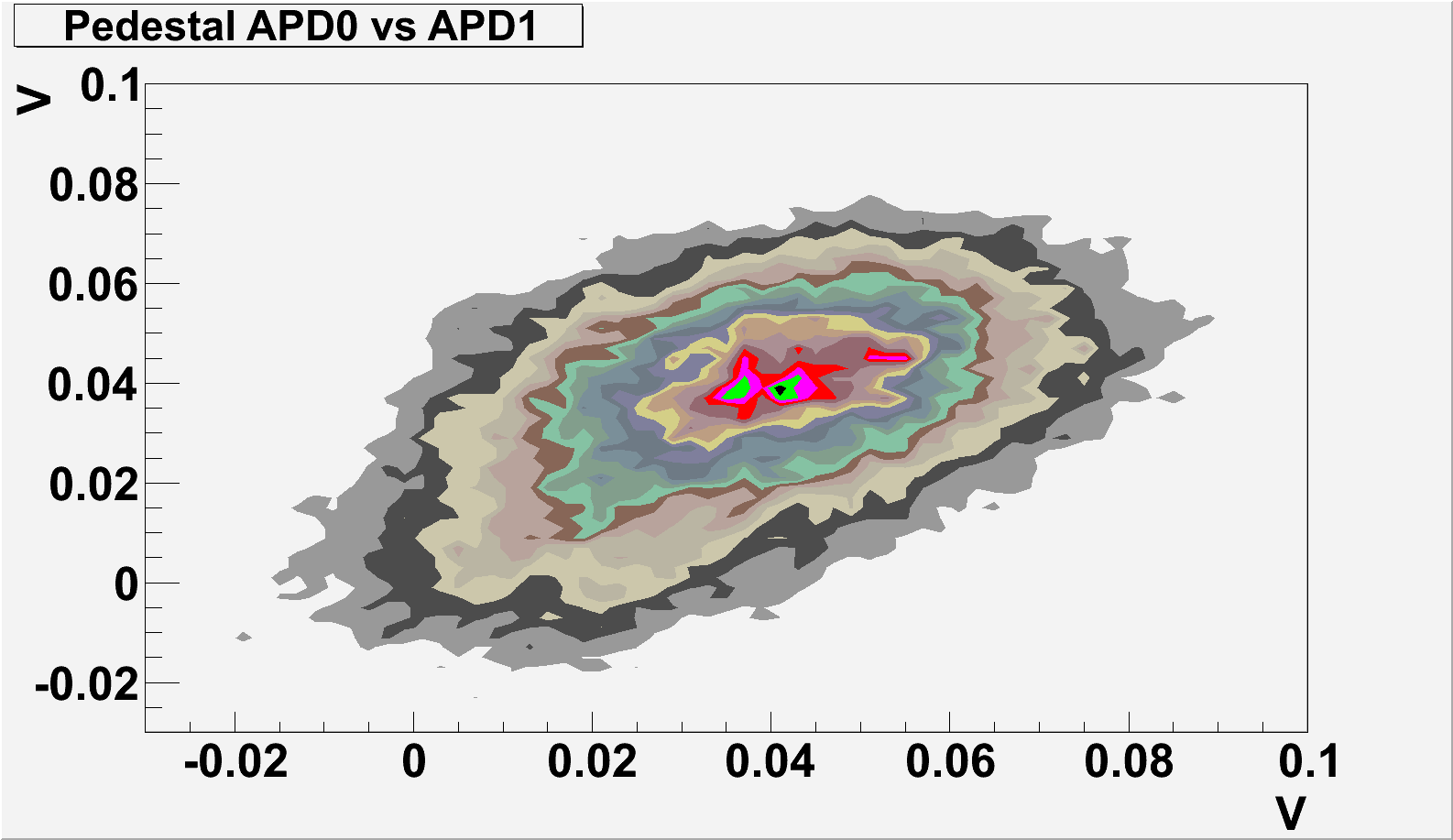}
\caption{\label{fig_common_noise}Pedestal of APD 1 versus pedestal of APD2 showing a high correlation ($\approx$70 \%) between them.}
\end{minipage}
\end{center}
\end{figure}

The steel chamber is a cylindrical pressure vessel with a diameter of 18 cm and a height of 8 cm. It has two CF16 flanges for the connection to the gas system. The gas system consists of a loop between the two flanges which includes a hot getter (SAES ST707) for purification and acts at the same time as recirculation pump due to convection. The interior contains a tower made of PEEK. It consists of a copper coated Kapton cathode, two EL meshes (wire diameter: 80 $\mu$m, wire pitch: 900 $\mu$m) and the holder for the 5 APDs.
The drift distance between the cathode and the upper EL mesh is 15 mm, the gap between the meshes is 7 mm and the distance between lower mesh and APD surface about 5 mm. Horizontally the sensitive volume has the size of about 30x30 mm$^2$. The signal is carried out by 9-pin Lemo feedthroughs while the Macor HV connectors were built at IFAE. To allow X-rays to enter the chamber a Kapton window is located in the center of the end cap behind the cathode. Due to the distance of about 15 mm between the window and the cathode, low energetic X-ray sources cannot be used with this setup. The setup is shown in fig.~\ref{fig_chamber1},~\ref{fig_chamber2} and ~\ref{fig_chamber3}.

\subsection{The APDs}
For the readout of the chamber, 5 APDs (S8664-55-SPL) from Hamamatsu were used. These are standard APDs without the protection window which normally absorbs VUV photons. The APDs have a size of 5x5 mm$^2$%, a 10x10 mm$^2$ version is also available, 
and were placed with a pitch of 10.3 mm between them. These APDs are directly sensitive to the EL light of xenon (172 nm) with a quantum efficiency of about 80 \% accoring to the data sheet. Measurements performed together with the research group at the University of Coimbra gave a result of $70\pm16 \%$. They are supposed to have also a high quantum efficiency to argon light (128 nm) which could amplify their field of application even more. Additional studies in respect of this are planned. The same bias voltage is applied to all 5 APDs, while the signal from each is read out independently. This powering scheme introduces a high correlation between the noise of the APDs. Fig.~\ref{fig_common_noise} shows this for two APDs. Proper treatment of the pedestal correlation might help improve further the energy resolution and it is under investigation.

\subsection{Electronics and DAQ}
While for single sensor setups commercial electronics can be used, the development of a multi-sensor readout requieres the design of custom made electronics. 
The APD readout electronics developed at our institute is based on a transimpedance amplifier which converts the output current at the anode of the APD to voltage. This first stage of the readout has two main goals, convert the current signal to voltage and amplify it in a single step. This amplifier has the highest gain of about 10$^5$ which does not reduce the amplifier bandwidth below the signal bandwidth yielding to the full collection of the signal. Following the first amplifier stage there is a semi-Gaussian shaper  which filters the signal eliminating both high and low frequency noises. Afterwards the signal is digitized by a comercial ADC. The DAQ system to read the digitized data from the ADC and to store it, is based on the DAQ framework MIDAS \cite{Midas2010}. 

\section{Data taking and analysis} 
For the signal creation, a $^{109}$Cd source was used which emits mainly X-rays at 22.4 keV and 25 keV. The source is located on top of the central APD. The data taking is triggered by a threshold requirement on the energy deposition in the central APD while no requirements are applied to the outer APDs. After 1000 data events, 10 events with random trigger are taken to determine the pedestal and noise continuously during the data taking. The gain and energy resolution were measured for a wide range of parameters as drift field, EL field, APD voltages at various pressures between 1 and 1.65 bar.  

\subsection{Intercalibration}
\begin{figure}[h]
\begin{center}
\begin{minipage}{14pc}
\includegraphics[width=14pc]{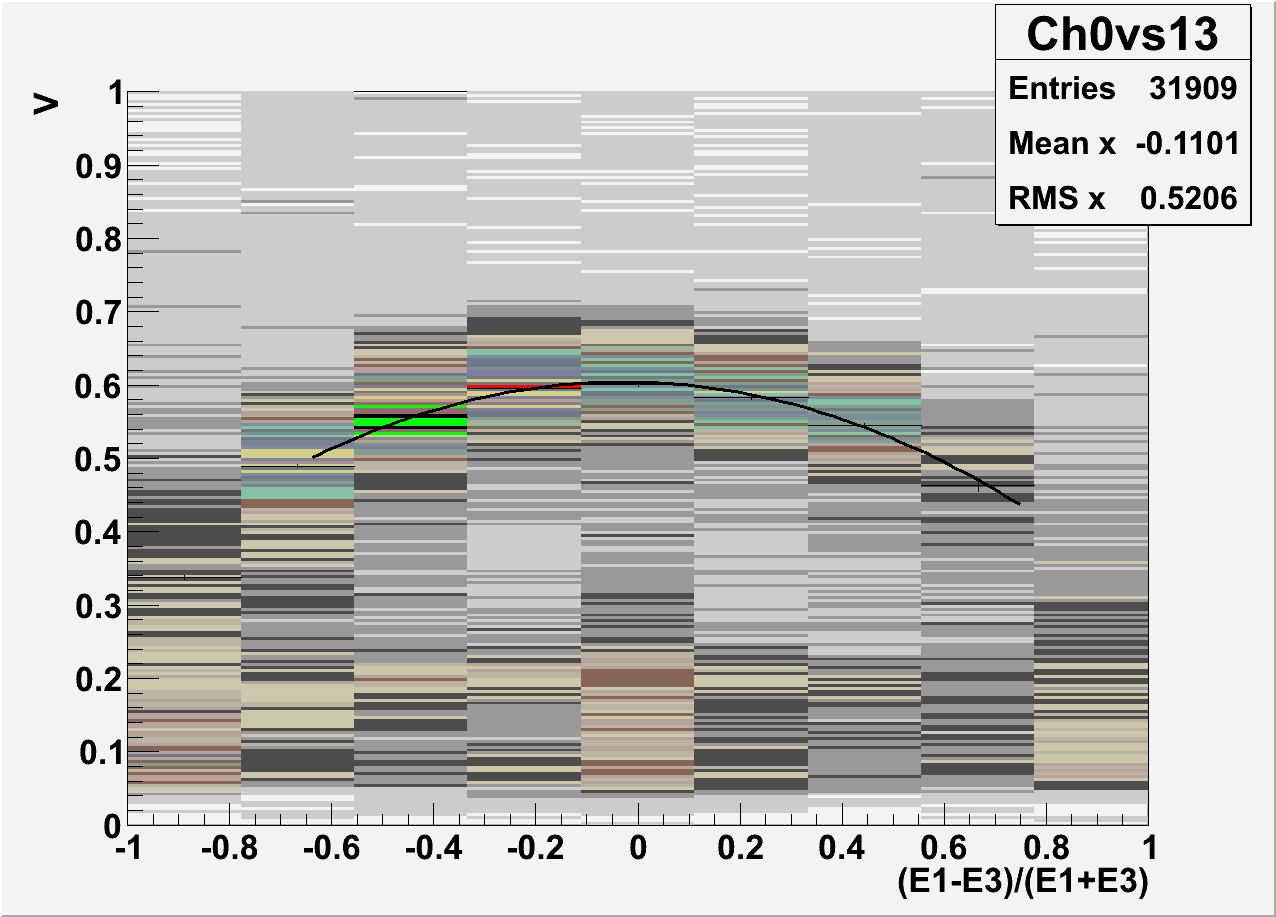}
\caption{\label{fig_intercal1}Plot of the energy in the central APD versus the asymmetry between APD 1 and 3.}
\end{minipage} \hspace{2pc}%
\begin{minipage}{16pc}
\includegraphics[width=16pc]{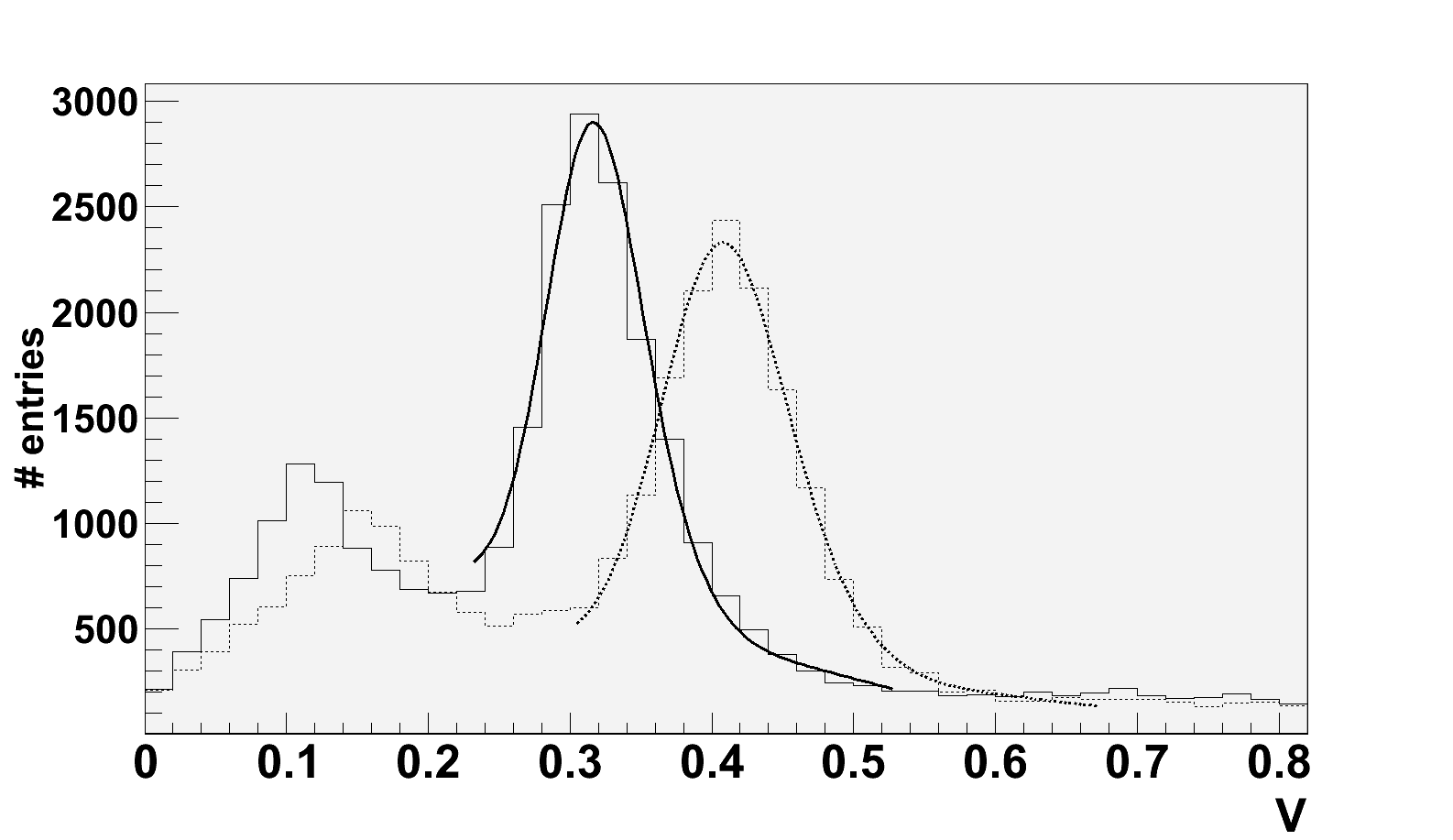}
\caption{\label{fig_intercal2}The spectrum of APD 1 and 3 (solid line) and of APD 2 and 4 (dotted line).}
\end{minipage}
\end{center}
\end{figure}

The operation of several APDs requires a careful intercalibration among them. The intercalibration was performed in various steps. In a first step, the two outer APDs on one axis were intercalibrated from the plot of the asymmetry, defined as $(E_i-E_{i+2})/(E_i+E_{i+2})$, between them versus the energy deposition in the central APD as shown in fig.~\ref{fig_intercal1}. If no intercalibration is required the curve fitted to this distribution will have its maximum at 0, otherwise the observed shift from 0 can be used for the intercalibration. In the next step, the two axis were intercalibrated using the $^{109}$Cd spectrum obtained with the outer APDs on each axis (see fig.~\ref{fig_intercal2}). In the final step, the central APD is intercalibrated with the outer ones by minimizing the width of the 22 keV peak of the spectrum of all 5 APDs obtained from $E=E_0+\alpha E_{out}$. The calibration factors depend on the gain of the APDs and therefore on the APD bias voltage but also external factors as the temperature\footnote{not monitored until now}. However, for the moment, as long as not explicitly mentioned, the calibration factors for an APD voltage of 400 V are applied to all data.   

\subsection{Position determination and first tracks}
\begin{figure}[h]
\begin{center}
\begin{minipage}{14pc}
\includegraphics[width=14pc]{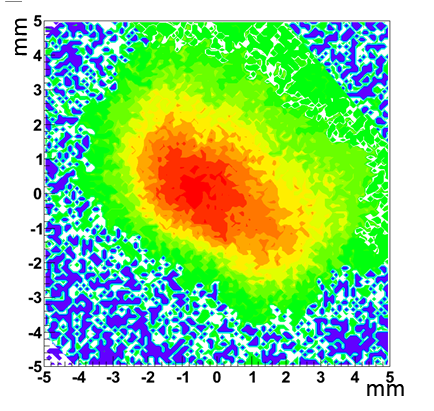}
\caption{\label{fig_xyplot}The $xy$ distribution of the X-ray conversions. Most of the conversions happen in the central region (red area).}
\end{minipage} \hspace{2pc}%
\begin{minipage}{14pc}
\includegraphics[width=14pc]{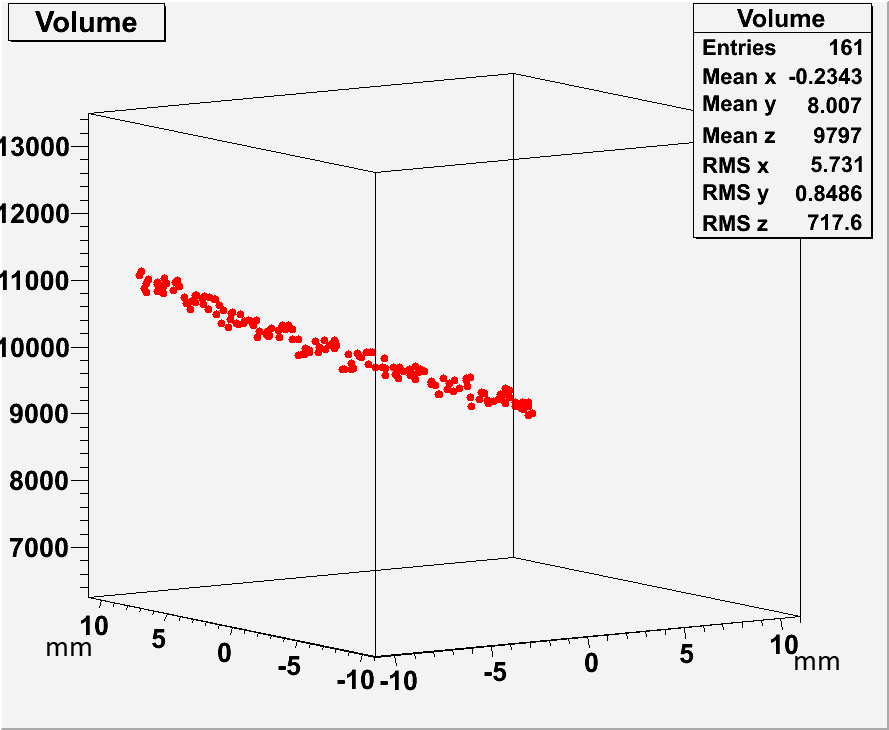}
\caption{\label{fig_track} A track crossing the active volume under a large angle in respect of the $z$-axis.}
\end{minipage}
\end{center}
\end{figure}

The position of the X-ray conversion in the $xy$-plane is determined by the center-of-gravity method.
%\begin{equation}
%x/y=\frac{\sum{C_iA_ix/y_i}}{\sum{C_iA_i}}
%\end{equation}
%Here, $C_i$ is the calibration factor of APD $i$, $A_i$ the charge deposition in that APD and $x/y$ its $x$ respectively $y$ position.
The spatial distribution is shown in fig.~\ref{fig_xyplot}. Most of the conversions happen within the central region. The deviation from the expected rotation-symmetric distribution might be caused by a slight misalignment between the source and the center of the central APD or by imperfections in the used collimator.\\
Cosmic runs without source are used for first tracking attempts. The center-of-gravity of the charge distribution is calculated for each time bin. Fig.~\ref{fig_track} shows one of the tracks reconstructed in this way. A point like charge deposition would produce only a straight line along the $z$-axis.

\subsection{Energy resolution and threshold}
\begin{figure}[h]
\begin{center}
\begin{minipage}{14pc}
\includegraphics[width=14pc]{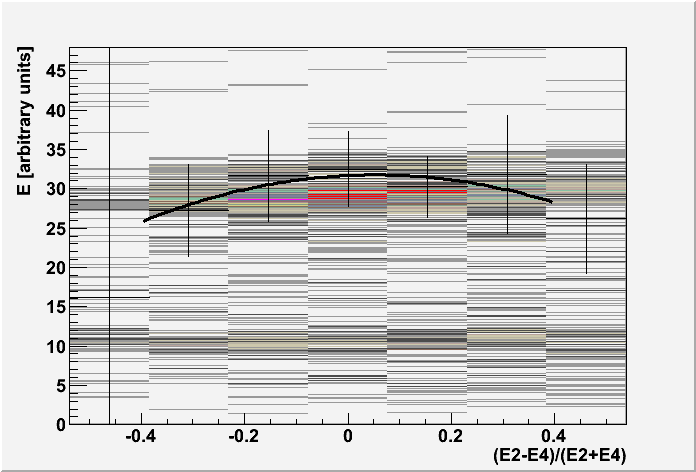}
\end{minipage}
\begin{minipage}{14pc}
\caption{\label{fig_totEasym}The asymmetry between APD 2 and 4 versus the summed energy over all APDs. The same was done for the asymmetry between APD 1 and 3.}
\end{minipage} 
\end{center}
\end{figure}

The energy is measured with the following method:
\begin{itemize}
\item Determining the time bin position of the maximum of the central APD.
\item Calculating the average charge deposition per voxel in a range of $\pm3 \mu$s around the maximum for each APD.
\item Subtracting the pedestal for each APD.
\item Summing over all APDs.
\item Correcting for the total energy depending on the up-down/left-right asymmetry.
\end{itemize}
Fig.~\ref{fig_totEasym} shows the total energy deposition as function of the asymmetry. One clearly sees a quadratic dependency caused by the smaller acceptance towards the edges.

\begin{figure}[h]
\begin{center}
\begin{minipage}{14pc}
\includegraphics[width=14pc]{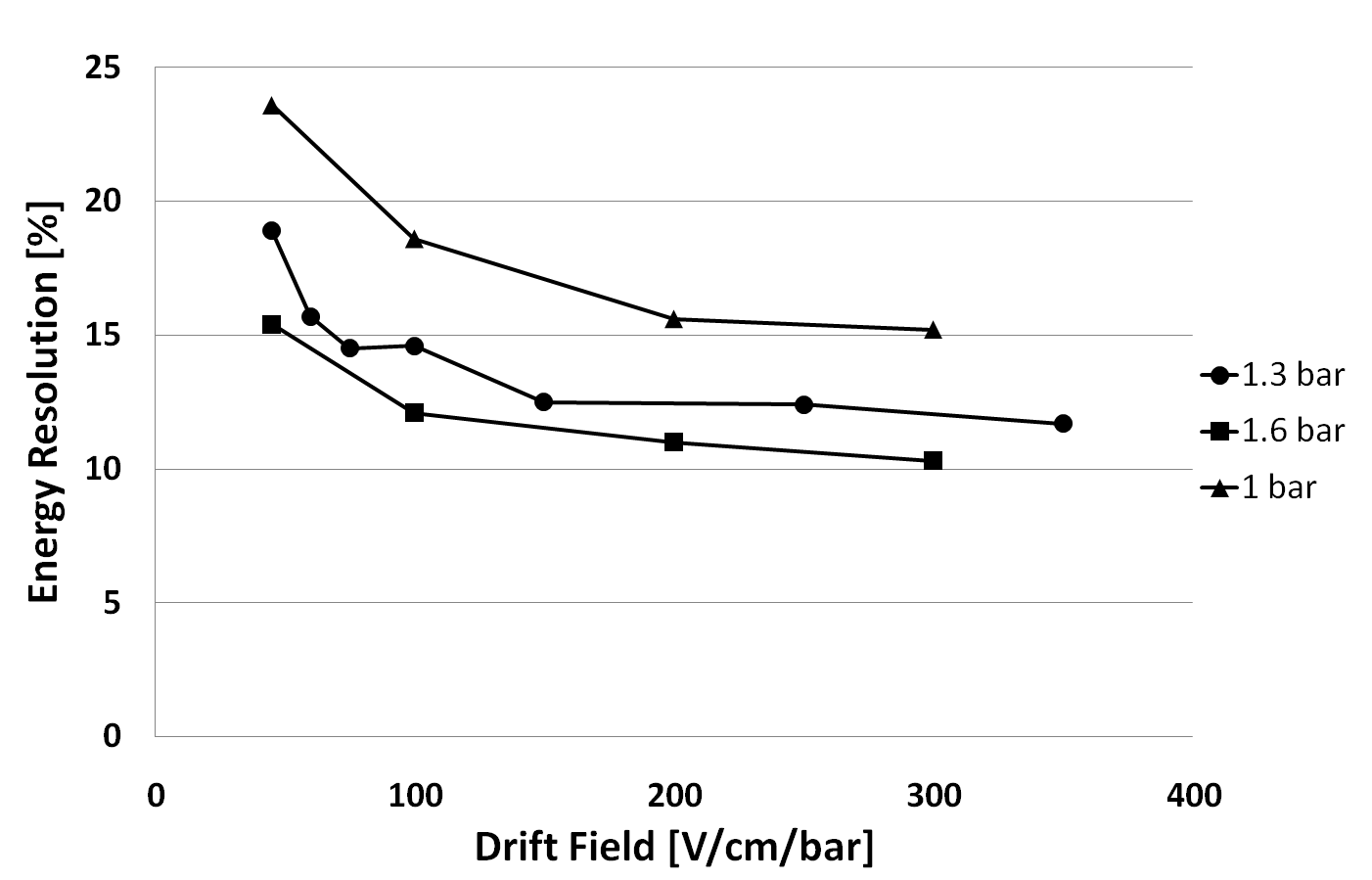}
\caption{\label{fig_scan_drift}Scan of the drift field. The settings for the other parameters were fixed for this measurement series (EL field: 3.5 kV/cm/bar, APD bias voltage: 405 V). }
\end{minipage}\hspace{2pc}%
%\end{center}
%\end{figure}
%\begin{figure}[h]
%\begin{center}
\begin{minipage}{14pc}
\includegraphics[width=14pc]{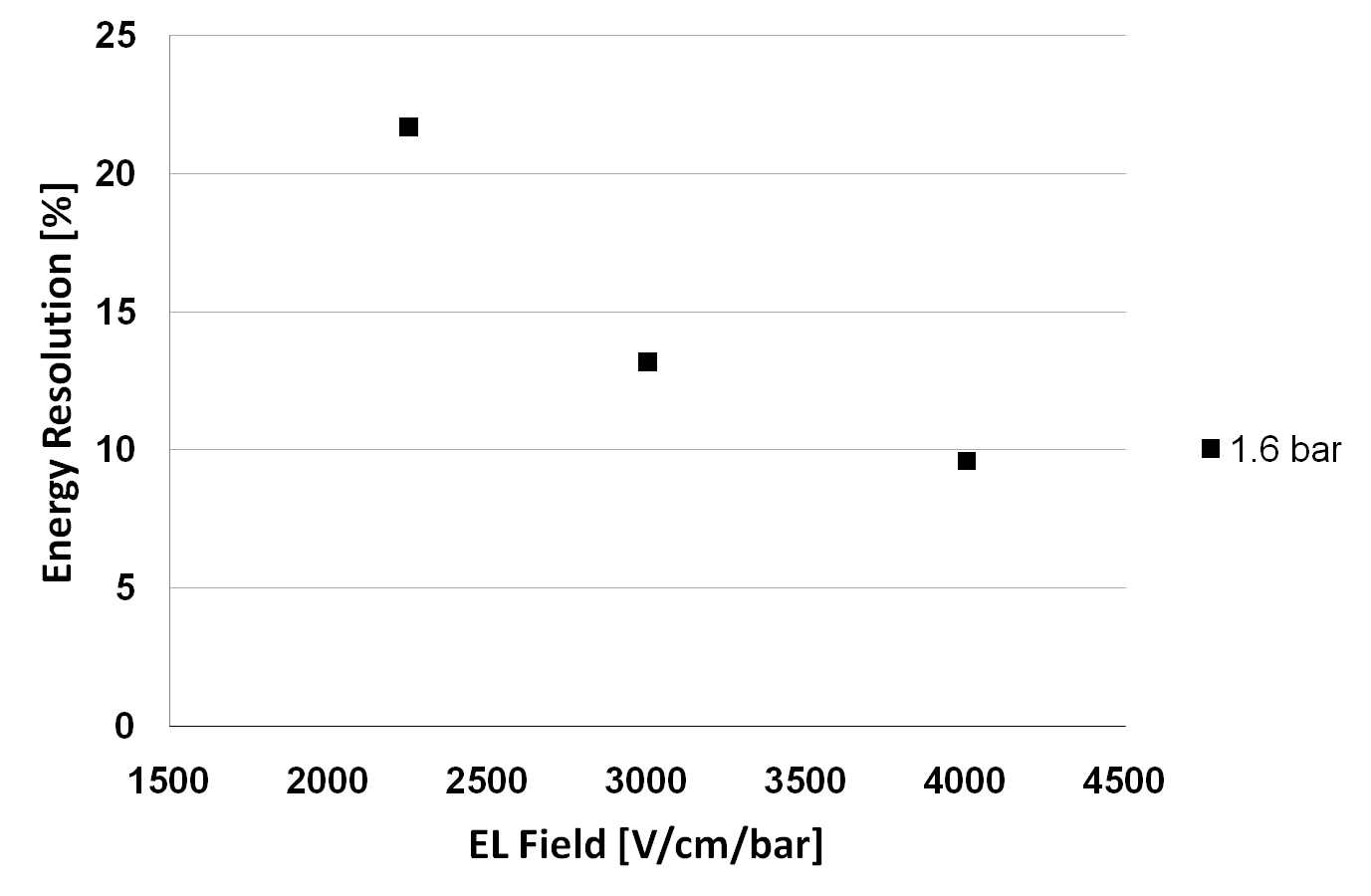}
\caption{\label{fig_scan_EL}Scan of the EL field. The settings for the other parameters were fixed for this measurement series (drift field field: 200 V/cm/bar, APD bias voltage: 405 V).}
\end{minipage}\hspace{2pc}%
\begin{minipage}{14pc}
\includegraphics[width=14pc]{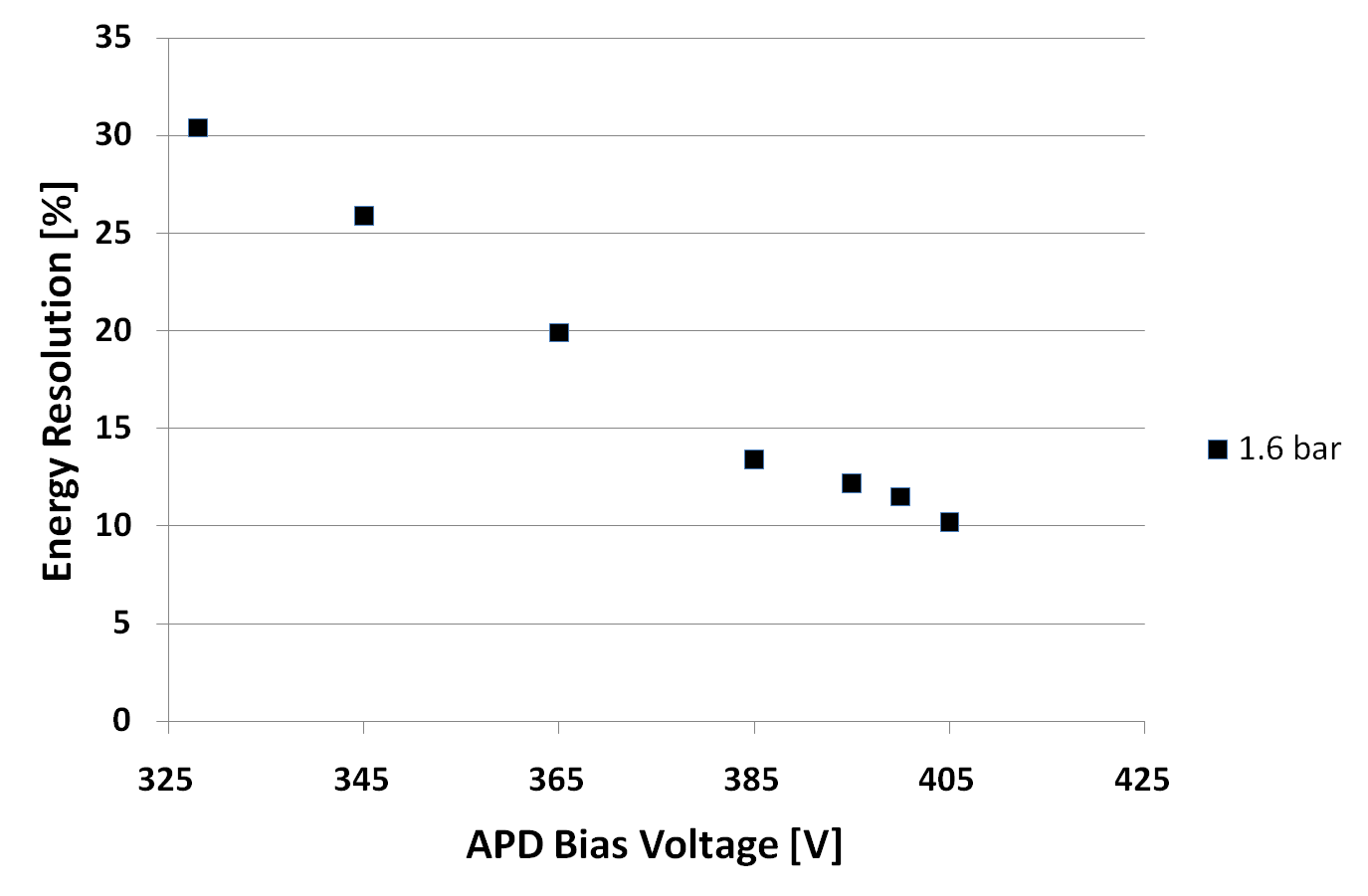}
\end{minipage} \hspace{2pc}
\begin{minipage}{14pc}
\caption{\label{fig_scan_APD}Scan of the bias voltage. The settings for the other parameters were fixed for this measurement series (EL field: 3.5 kV/cm/bar, drift field: 200 V/cm/bar).}
\end{minipage} 
\end{center}
\end{figure}

\begin{figure}[h]
\begin{center}
\begin{minipage}{24pc}
\includegraphics[width=24pc]{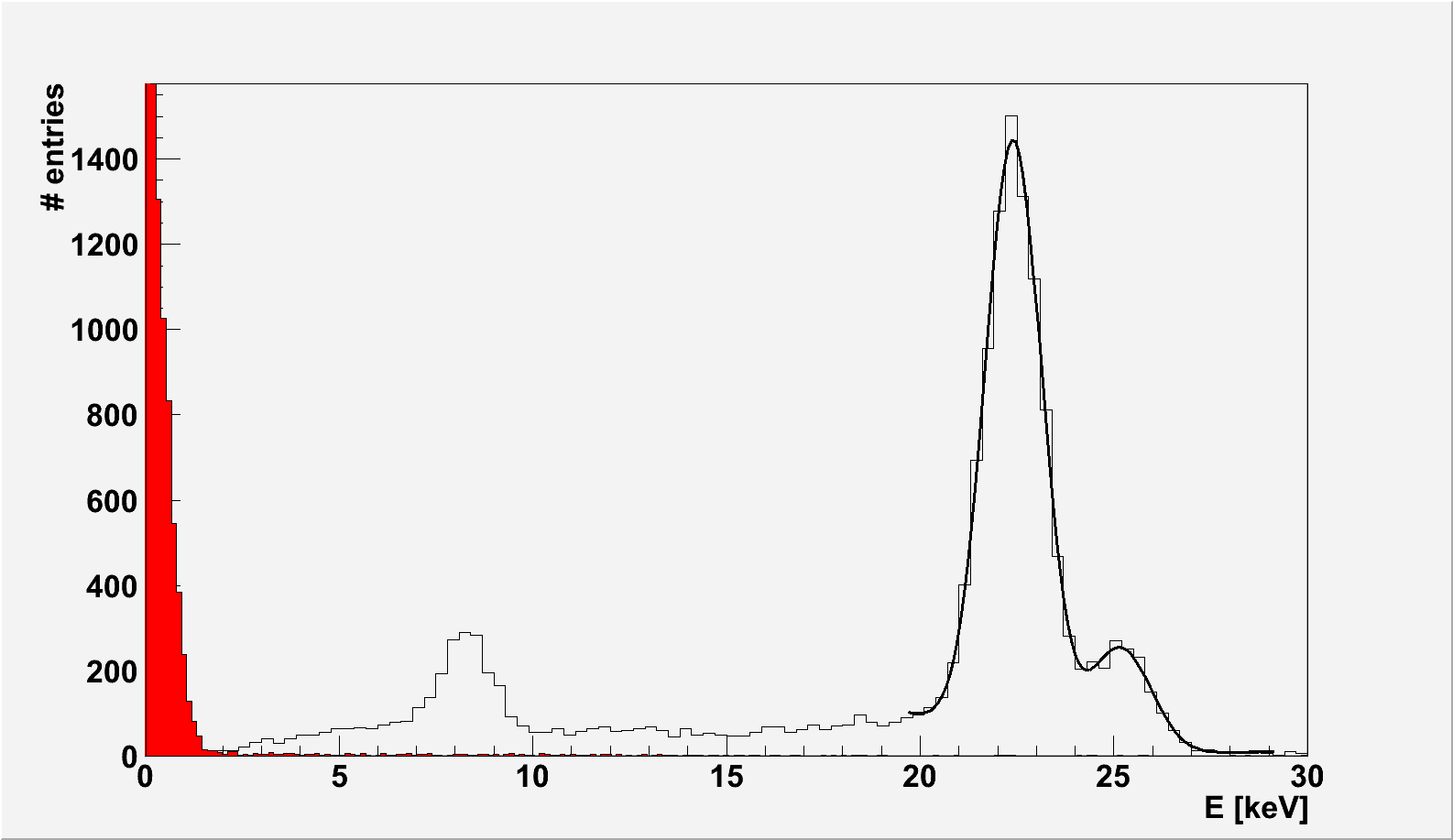}
\caption{\label{fig_bestspectra}The best result achieved for the energy resolution was: $(7.7\pm0.1)\%$ FWHM for the 22.4 keV line. The settings for this run were: drift field: 300 V/cm/bar, EL field: 4 kV/cm/bar, APD bias voltage: 410 V, pressure 1.65 bar. }
\end{minipage}\hspace{2pc}%
\end{center}
\end{figure}

\begin{figure}[h]
\begin{center}
\begin{minipage}{14pc}
\includegraphics[width=14pc]{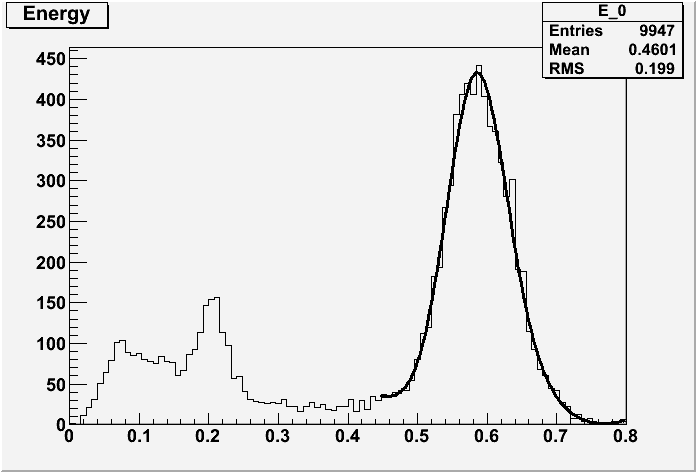}
\caption{\label{fig_extreme1} The spectrum of the central APD alone. No corrections were applied to the data. The energy resolution in this case is 17.5 \%.}
\end{minipage}\hspace{2pc}%
\begin{minipage}{14pc}
\includegraphics[width=14pc]{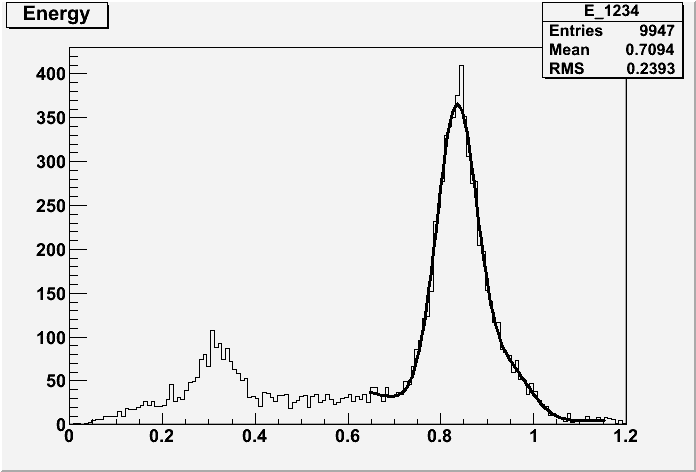}
\caption{\label{fig_extreme2} The spectrum of the outer APDs. Also here, no corrections were applied to the data and an energy resolution of 12.4 \% was achieved.}
\end{minipage} 
\end{center}
\end{figure}

First systematic scans of the energy resolution as function of various parameters were performed to find the optimal settings. Fig.~\ref{fig_scan_drift} shows the scan of energy resolution as function of the drift field for 3 different pressures. The energy resolution saturates for drift fields above 200 V/cm/bar for all pressures. As expected, it was found that the result improves with increasing pressure. Fig.~\ref{fig_scan_EL} shows the energy resolution as function of the applied EL field. The energy resolution improves significantly with higher EL fields with potential for further improvements of the results by running at fields higher than 4 kV/cm/bar. For the moment, the field is limited to this value to avoid HV instabilities within the chamber. The scan of the APD bias voltage is shown in fig.~\ref{fig_scan_APD}. The curve has a steep falling shape for low APD voltages  and starts to saturate at a voltage of about 390 V. For this scan, the maximal APD bias voltage was conservatively chosen to 405 V. However, a further improvement of the energy resolution might be achieved by exploiting the full phase space of the voltages up to 425 V.\\    
The best result for the energy resolution of $(7.7\pm 0.1)$ \% FWHM is shown in fig.~\ref{fig_bestspectra}. The data are restricted to an area of $\pm 5$ mm around the center of the chamber. The calibration factors applied correspond to the applied voltage of 410 V to the APDs. The normalization is chosen such that the main peak corresponds to the literature value of 22.4 keV for the main line of $^{109}$Cd. Also the peak of 25 keV is clearly visible. The additional peak which can be found at 8 keV is caused by X-ray fluorescence of the copper of the cathode. This peak has an energy resolution of 19.1 \% which is worse than expected when scaled with $\sqrt{E}$, we assume due to a higher noise contribution. The red area shows the "`signal"' created by the pedestal when the same method as for the data is applied to a random time interval of 6 $\mu$s. One can conclude that a low threshold of about 2 keV can be achieved with such a readout at 1.65 bar.
In respect of the achievable energy resolution, it is also interesting to investigate the two extreme cases: The energy resolution achievable when only the central APD is used for the energy measurement and the case when only the outer APDs are used. The latter case is of interest to study the dependency of the energy resolution on the pitch between the APDs when the conversion occurs within the gap between the APDs. The results are shown in fig.~\ref{fig_extreme1} and fig.~\ref{fig_extreme2} for the same data sample as before. 
%For these plots no corrections were applied to the data. Nevertheless 
Energy resolutions of 17.6 \% FWHM for the central APD and 12.4 \% FWHM for the outer APDs were achieved. The latter value suggests that even with a larger pitch an energy resolution of $\leq 1$ \% at 2.46 MeV could be reached.       
\section{Conclusion and outlook}
A 5 APD readout system was successfully commissioned and operated for the first time to readout a small EL detector filled with xenon at pressure of up to 1.65 bar. 
Very good energy resolutions of 7.7 \% FWHM at 1.65 bar were achieved for 22.4 keV X-rays. This corresponds to about 0.7 \% FWHM at 2.46 MeV. Further improvement of this result seems to be likely by operation at higher pressure (see fig.~\ref{fig_coimbra1}), further optimization of the APD bias voltage and the EL field and temperature corrections. It was shown that such a readout could also provide a very low detection threshold of a few keV. In addition position determination and first tracking were performed.\\
For the future it is planned to extend the existing system of 5 APDs to 30 APDs in a chamber with an active volume of 20 cm diameter and 11 cm length. The system is currently under construction and should give more information about the possibilities of such a readout system for EL detectors.

\ack
The authors acknowledge the support received by the CONSOLIDER INGENIO Project CSD2008-0037 (CUP) and the FPA2009-13697-C04-03 project.
We also want to thank Javier Gaweda for his technical support. 

\section*{References}
\bibliography{lib}
\bibliographystyle{iopart-num}

\end{document}